\documentstyle[eqsecnum,pre,preprint,aps,epsfig]{revtex}

\tightenlines

\newcommand{\bq}{\begin{equation}}
\newcommand{\eq}{\end{equation}}
\newcommand{\bqa}{\begin{eqnarray}}
\newcommand{\eqa}{\end{eqnarray}}
\newcommand{\ben}{\begin{enumerate}}
\newcommand{\een}{\end{enumerate}}
\newcommand{\bc}{\begin{center}}
\newcommand{\ec}{\end{center}}
\newcommand{\bqb}{\begin{eqnarray*}}
\newcommand{\eqb}{\end{eqnarray*}}

\newcommand{\gaugelog}{\log\frac{s}{M_W^2}}
\newcommand{\gaugeloglog}{\log^2\frac{s}{M_W^2}}

\begin{document}

\draft
\preprint{PM/04-58,~~December, 2004,~~
hep-ph/04......}

\title{\vspace{1cm}  Split Supersymmetry 
at the Logarithmic Test of Future Colliders.
\footnote{Partially supported by EU contract HPRN-CT-2000-00149}}
\author{M. Beccaria$^{a,b}$,
F.M. Renard$^c$ and C. Verzegnassi$^{d, e}$ \\
\vspace{0.4cm}
}

\address{
S$^a$Dipartimento di Fisica, Universit\`a di
Lecce \\
Via Arnesano, 73100 Lecce, Italy.\\
\vspace{0.2cm}
$^b$INFN, Sezione di Lecce\\
\vspace{0.2cm}
$^c$ Physique
Math\'{e}matique et Th\'{e}orique, UMR 5825\\
Universit\'{e} Montpellier
II,  F-34095 Montpellier Cedex 5.\hspace{2.2cm}\\
\vspace{0.2cm}
$^d$
Dipartimento di Fisica Teorica, Universit\`a di Trieste, \\
Strada Costiera
 14, Miramare (Trieste) \\
\vspace{0.2cm}
$^e$ INFN, Sezione di Trieste\\
}

\maketitle

\begin{abstract}

We consider a large number of pair production processes at future
colliders (LHC, ILC) for values of the c.m. energy in the TeV range,
where a logarithmic expansion of Sudakov kind would provide a reliable
description of Split supersymmetric electroweak effects.
We calculate all the leading and next to leading terms of the
expansions, that would differ drastically
in the considered domain from those of an extreme "light" scenario. 
We imagine then two possible competitive future situations, at LHC 
and at ILC,
where the determination of the energy dependence of the cross
sections of certain processes could reveal a "signal"
of the correct supersymmetric scheme.

\end{abstract}
\pacs{PACS numbers: 12.15.-y, 12.15.Lk, 13.75.Cs, 14.80.Ly}

\section{Introduction}

The theoretical picture of no low scale Supersymmetry 
\cite{AD} or Split Supersymmetry \cite{GR} 
has been proposed very recently 
as a possible solution to 
several problems that still remain in the original MSSM formulation, 
essentially based on the expectation of a "TeV Supersymmetry" 
with all 
sparticles' masses roughly of the order of 1 TeV or less. 
Given the relevance 
of the proposal, a number of authors \cite{ADGR,Zhu,KPRC} have 
already suggested that 
indirect searches of signals of the model might become 
available in a not too 
far future, also via precise measurements to be 
carried on at the Large Hadron Collider (LHC) and, 
eventually, at the planned future International Linear Collider (ILC). 
In particular, the production 
of chargino and neutralino 
pairs has been indicated as a good indicator of 
Split Supersymmetry effects, 
with the feeling, though, that the requested experimental 
accuracy for the 
processes might be more realistically obtainable at ILC.\par
Generally speaking, the search of low energy signals
of any model is based on 
the fact that there exist effects that are "visibly" 
different from those of 
other competitors. In the case that we are considering, 
one might view the two 
proposals of "TeV Supersymmetry" and of "Split Supersymmetry" 
as the models to 
be examined. To make this examination meaningful, 
one should also assume a 
situation where, briefly, at least one of the two 
Supersymmetries will be the 
correct one. This initial attitude will be the 
starting point of our paper. In 
other words, we shall postulate that some of the light supersymmetric 
components that the two models share, in practice charginos and/or 
neutralinos, have been discovered (together with the 
light neutral Higgs boson 
of the theory), and that no indication of the 
remaining scalar supersymmetric 
particles is still available. 
In this spirit, we shall investigate the 
possibility of measuring a simple and clean property of 
Split Supersymmetry 
that might be considered as a possible signal to be 
taken into account.\par 
The bulk of our analysis will be the observation 
that in a supersymmetric model
there exists, for a general process of pair production 
due to the annihilation 
of an initial elementary (electron-positron or parton-parton) pair, 
a range of 
the initial pair c.m. energy where a simple type 
of logarithmic expansion of 
so called Sudakov kind could be used to describe 
"Split Supersymmetry " electroweak
one-loop effects, but not "TeV Supersymmetry" ones. 
This would correspond to 
values of the initial c.m. energy in the one TeV range, realistically 
achievable both at LHC and at ILC.\par 
The purpose of this preliminary paper is 
that of computing and of listing the aforementioned Sudakov 
Split expansions 
for all those processes of pair production that would 
be common to the two 
supersymmetric schemes, and measured either at LHC or at ILC. 
These new 
results will be displayed in the following Section 2, and
compared for sake of completeness with the corresponding 
expansions already 
obtained by us in previous papers for the case that was called of 
"moderately light" Supersymmetry, where all SUSY masses 
were supposed to lie 
below, roughly, 400 GeV, the main motivation of this 
(academic) comparison 
being that of showing that the two expansions would be 
drastically different. 
Having at disposal the complete list of "Split Supersymmetry" Sudakov 
expansions, we shall try in Section 3 to provide two special 
cases of 
applications of our results as a possible way of differentiating "TeV 
Supersymmetry" from "Split Supersymmetry" in realistic experimental 
situations, keeping in mind that, while the relevance 
of the examples remains 
completely hypothetical, the validity of the Sudakov 
expansions will be,conversely, true. A short final discussion, given in Section 4,
will conclude the paper.

\section{Electroweak Sudakov expansions for Split Supersymmetry}

In order to follow a reasonably sequential chronological scheme,
we shall 
first consider in Section 2A the case of pair production at LHC, 
having in mind the
reasons that will select the initial parton pair c.m. energy 1 TeV 
range for our analysis. Section 2B will be devoted to the analogous 
investigation of pair production at ILC, for a similar energy choice.

\subsection{Pair production from an initial parton-parton state at LHC}

Independently of the nature of the assumed supersymmetric model, 
LHC will certainly measure the production of Standard Model pairs. 
Within this area, special interest has been devoted to the processes 
of top-antitop production and of single top 
(i.e. $td$, $t\bar b$, $tW^-$) 
production \cite{CERNYB} for which the estimate of 
realistic experimental and theoretical uncertainties has been, and is being, actively 
performed. This motivates our choice of beginning our 
analysis with the determination of the proper Split Supersymmetry 
logarithmic expansions for these processes. 
Actually, we already performed a similar calculation in previous 
papers~\cite{quark,ttb}, assuming the validity of what 
was called a "moderately" light SUSY scenario, 
with all sparticle masses lighter than, approximately, 400 GeV. 
In such a situation, we concentrated our study on 
initial partonic pair c.m. energy $\sqrt{s}$ values 
varying in the 1 TeV  region, since we expected the validity,
in that range, of a logarithmic Sudakov expansion 
of the scattering amplitude whose coefficients were known 
to next-to leading (i.e. linear) logarithmic order. 
This expectation was a consequence of the 
"technical" fact that the c.m. $\simeq 1$ TeV energy was sufficiently 
larger than all the MSSM particle-sparticle masses 
involved in the processes. 
In particular, at the chosen one-loop perturbative 
order, the relevant sparticles were charginos, neutralinos, 
squarks, charged Higgs bosons (for the electroweak effects) and gluinos 
(for the strong ones).  
As stressed in that Reference, a welcome consequence 
of that energy choice was also the fact that, in the 1 
TeV c.m. energy range, several kinematical 
simplifications were arising, allowing to neglect 
a number of contributions to the processes. 
In practice, given the nature of the investigated SUSY effects
that were systematically of higher order, it turned out that 
for a meaningful analysis it was sufficient to consider  
the one-loop corrections to the Born $t$ and $u$ channels
exchange processes for top-antitop 
production from an initial gluon-gluon state, 
and to treat the quark-antiquark 
contribution for this process in 
Born approximation. Remarkable simplifications were also valid 
for ($t,W^-$) production, and we defer the reader to Ref.\cite{quark} 
for more details.\par
The main result of the analysis of Ref.\cite{quark} was 
the fact that, at the chosen one-loop order, 
the supersymmetric electroweak effects computed, in the usual 
approximation, at "next-to-leading order" 
(i.e. retaining the quadratic and 
the linear logarithmic terms of the expansion) 
were systematically (for all the considered processes) "large", 
i.e. well beyond the relative ten percent size, 
particularly for large $\tan\beta = v_2/v_1$ values. 
QCD supersymmetric effects, conversely, turned out to be 
definitely smaller (of the few per cent relative size),
but of the same sign as 
the electroweak ones, thus increasing 
the overall predicted SUSY virtual contribution.
At this level, accurate dedicated experimental measurements of the 
processes should be able to "see" the effect, 
thus providing a relevant test of the model, as discussed e.g. in a 
very recent paper \cite{ttb}.\par
From a technical point of view, the origin of 
the "large" SUSY Sudakov  effects is mostly due to the 
contributions coming from the vertices where 
couplings of Yukawa type of the top quark appear. 
These involve the presence of virtual
combined gauginos-third family  squarks and SUSY 
Higgses-heavy quarks electroweak diagrams; SUSY QCD (SQCD) effects are 
only provided in the chosen situation by 
vertices with combined gluino-squark diagrams. 
There would also be (less relevant) SUSY logarithmic contributions of 
Renormalization Group origin, but only in the so called t-channel 
(final $td$) and s-channel (final $t\bar b$) 
single top production, coming e.g. from gauginos bubbles in the W 
propagator. We insist on the fact that all the 
mentioned terms can be estimated in the simple logarithmic 
expansion as a consequence of the fact that all the 
involved masses are supposed to be sufficiently smaller that the chosen 
$\simeq1$ TeV c.m. energy $\sqrt{s}$. 
It should be also stressed that, from an 
experimental point of view, for the chosen processes, 
this energy range is  statistically valid; actually, one 
could probably enlarge the range until "extreme" 
values of $\sqrt{s} \simeq1.5$ TeV, and 
still find a reasonable number of events to be considered
\cite{ttb}. Note also that, in the considered MSSM scheme, 
there appear  electroweak Sudakov terms coming from 
Standard Model virtual particles exchanges, 
in particular vertices or boxes with gauge 
bosons  and vertices with the SM Higgs boson. For all these particles
and for the top quark as well, the chosen value of 
$\simeq1$ TeV c.m. energy is clearly sufficiently high 
to guarantee the validity of the 
simple asymptotic Sudakov logarithmic expansion.\par
Starting from these observations, it seems now almost natural to us to 
continue to consider as a convenient energy range, 
in which to compute a simple expression of Split Supersymmetry effects 
for the aforementioned processes, the previously considered one i.e. 
$\sqrt{s}$ in the 1 TeV range. 
In fact, for such energy values, all the contributions 
from the electroweak SUSY vertices of Yukawa kind will, simply, 
decouple and disappear, since they involve either superheavy 
squarks or superheavy SUSY Higgses. 
The same will happen to the SUSY QCD gluino-squark vertices (in fact, 
they disappear for the same reason that would make the decays 
of the possibly light gluino of the model to be hardly detectable). 
Only the (small) RG SUSY effects due to pure gauginos 
exchanges (e.g. in the W propagator) will remain. 
Thus, the electroweak Sudakov logarithmic expansion (which, 
by definition, does not contain the RG terms) 
will be exactly the same as in the SM case, 
since those  contributions will remain unmodified. 
At $\sqrt{s}\simeq1$ TeV at LHC, 
for the chosen processes, one 
would find in conclusion, in case of Split Supersymmetry,  
the same logarithmic Sudakov expansion (to next-to leading order) 
given by the Standard Model!\par
After this long but, we hope, useful preliminary 
discussion, we are now ready to list the Sudakov expansion 
for Split Supersymmetry for a number of 
processes, starting with those that have been previously mentioned.
To these expansions we shall add (and list) the corresponding 
ones that were obtained \cite{quark,ttb,sud} 
in the case of "moderately light" 
Supersymmetry, essentially to evidentiate the big 
numerical difference between the effects in the two cases. 
We shall define in this paper this second scenario as 
"moderately light MSSM" and indicate it with the 
shorthand notation "m.l. MSSM". To this list, of 
mostly academic relevance, we shall finally add a 
third one that corresponds to 
a situation in which all the squarks are "relatively" heavy, 
in particular with a mass beyond the final LHC limit,
assumed to be of approximately 1.5-2 TeV \cite{KM} 
(reasonably different values would also be equivalent for our 
purposes), while the Supersymmetric (neutral and charged) 
Higgs bosons and sleptons are still "moderately" 
light, with a mass not much above the expected LHC reach,
assumed to be of, roughly, 400 GeV\cite{KM}.
We shall define this scenario, that will be numerically 
examined in the final part of 
this paper, "moderately light MSSM with heavy squarks", 
and indicate it as "h.s. MSSM". 
In our approach, for c.m. energy values in the 1 TeV range, 
the electroweak contributions coming from Feynman diagrams 
of Yukawa kind including Higgs 
bosons will still be correctly described in the ``h.s. MSSM'' by a 
logarithmic Sudakov expansion. 
This will not be a reliable attitude for the 
diagrams containing the "TeV scale" squarks. 
The latter would still contribute, although we would expect in 
a reduced way with respect to the light case as a consequence of the 
relatively large squark masses, 
but with an energy dependence that should be 
rather different from the logarithmic one of 
the "light" contributions. This dependence could be 
determined accurately by dedicated numerical calculations 
in the various cases. These are beyond the purposes of 
this preliminary work and will be investigated in details 
in a next paper. Keeping this limitation in mind,
we shall only quote in this Section the Sudakov contributions coming 
from the "moderately light" Higgs bosons in this third scenario.
For what concerns the notations and the conventions, 
we shall follow the same ones as in Ref.\cite{quark,ttb,sud}. 
To make, though, this Section more easily and quickly readable by 
a reader who were not too interested in the technical details, 
we have devoted a final Appendix to the complete definition 
of the several terms that appear in our formulae.\par
We are now ready to begin.
The processes that will be considered are the 
following ones:\\

{ \bf 1) \underline{$gg\to t\bar t$}}\\

For this process one has the 2 quantities ($\theta$ is the c.m. scattering angle and $s$, $t$, $u$ the
usual Mandelstam variables) :

\bqa
&&{\frac{d\sigma^{1~loop}_U}{
d\cos\theta} \stackrel{def}{=} \frac{d\sigma^{1~loop}(gg\to t_L\bar t_L +t_R\bar t_R)}
{d\cos\theta}}={d\sigma^{Born}(gg\to t_L\bar t_L +t_R\bar t_R)\over
d\cos\theta}(1+c^{t\bar t}_{L}+c^{t\bar t}_{R}) = \nonumber\\
&&={\pi\alpha^2_s\over4s}[ {u^2+t^2\over 3ut}-
{3(u^2+t^2)\over 4s^2}]~[1+{\alpha\over144\pi s^2_Wc^2_W}
(27-10s^2_W)(n_{qq}\log\frac{s}{M^2_W}
-\log^2\frac{s}{M^2_W})\nonumber\\
&&-~{\alpha \over16\pi s^2_W}\log\frac{s}{M^2_W}({3m^2_t\over M^2_W}
(\eta_Y+\eta'_Y \cot^2\beta)+{m^2_b\over M^2_W}
(\eta_Y+\eta'_Y \tan^2\beta))
-\eta{2\alpha_s\over3\pi}\log{s\over M^2}]
\label{sig1}\eqa

and the longitudinal polarization asymmetry 
\bq
a_t = \frac{\sigma(gg\to t_L\overline{t}_L)-\sigma(gg\to t_R\overline{t}_R)}{
\sigma(gg\to t_L\bar t_L +t_R\bar t_R)}
\eq
which reads (at one loop):
\bqa
a_t &\simeq&c^{t\bar t}_{L}-c^{t\bar
t}_{R}=
{\alpha(9-14s^2_W)\over48\pi s^2_Wc^2_W}
[n_{qq}\log\frac{s}{M^2_W}-\log^2\frac{s}{M^2_W}]
\nonumber\\
&&-{\alpha\over16\pi s^2_WM^2_W}\ \log\frac{s}{M^2_W}\
[m^2_b(\eta_Y+\eta'_Y \tan^2\beta)
-m^2_t(\eta_Y+\eta'_Y \cot^2\beta)]
\label{at1}\eqa
where $c^{t\bar t}_{L,R}=c^{ew}(t\bar t)_{L,R}+c^{SQCD}(q\bar q)$ 
are the coefficients defined in
Appendix A giving the one loop corrections;
 $s_W$ and $c_W$ are the sine and cosine of Weinberg angle, for which 
the LEP1 definition can be used, and:\\

in Split (as in SM):
$n_{qq}=3$, $\eta_Y=1$, $\eta'_Y=0$, $\eta=0$.\\ 

in "m.l. MSSM": $n_{qq}=2$, $\eta_Y=\eta'_Y=2$, $\eta=1$\\

in "h.s. MSSM": $n_{qq}=3$, $\eta_Y=1$, $\eta'_Y=1$, $\eta=0$.\\

Note that for this subprocess there is no RG term that would
differentiate the Standard Model (SM) from Split.\\

{ \bf 2) \underline{$b\ g\to t\ W^-$}}\\

\bqa
{d\sigma\over d\cos\theta}&=&
-~{\pi \alpha\alpha_s\over24s^2_Wus^2}
\{(s^2+u^2)
[1+c^{ew}(b\bar b)_L
+c^{ew}(t\bar t)_L+2c^{ew}(W^-_T)
\nonumber\\
&&+2c^{SQCD}(q\bar q)+2c^{ang,T}]
+{m^2_tt^2\over2M^2_W}[1+c^{ew}(b\bar b)_L
+c^{ew}(t\bar t)_R+2c^{ew}(W^-_0)
\nonumber\\
&&+2c^{SQCD}(q\bar q)+2c^{ang,0}]~\}
\eqa

The  universal quark coefficients (similar to the ones appearing in
the preceding process) are given explicitly for each model in
the Appendix. One needs also the following additional ones for
transverse and for longitudinal $W^-$:

\bq
c^{ew}(W^-_T)=~{\alpha\over 4\pi s^2_W}[-\log^2{s\over M^2_W}]
\eq

\bqa
c^{ang,T} = -~{\alpha\over2\pi}\
\log\frac{s}{M_W^2}\ \{~\log{-t\over s}\
{1-10c^2_W\over36s^2_Wc^2_W}
+~{1\over s^2_W}\log{-u\over s}~\}
\eqa

\bqa
c^{ang,0}=-~{\alpha\over24\pi c^2_W} \ \log\frac{s}{M_W^2}\
\{\frac{4}{3} \log{-t\over s}-{1-10c^2_W\over s^2_W}\log{-u\over s}~\}
\eqa
\noindent
(for any of the three considered model) and

\bqa
c^{ew}(W^-_0)
&=&{\alpha\over\pi}\ {1+2c^2_W\over32s^2_Wc^2_W}\ 
[n_{G}~\gaugelog-\gaugeloglog]
\eqa

where, in SM, Split and ``h.s. MSSM''  $n_{G}=4$, and
in "m.l. MSSM" $n_{G}=0$.\\

{ \bf 3) \underline{$bu\to td$ with $W$ exchange in the $t$ channel}}

\bqa
{d\sigma\over d\cos\theta}&=&
{\pi\alpha^2 s\over8s^4_W(t-M^2_W)^2}
[1+c^{ew}(b\bar b)_L+c^{ew}(u\bar u)_L+c^{ew}(d\bar d)_L
+c^{ew}(t\bar t)_L
\nonumber\\
&&+4c^{SQCD}(q\bar q)+2c^{ang}+2c^{RG}]
\eqa

The new specific coefficients are now:

\bqa
c^{ang}&=&-~{\alpha(1+8c^2_W)\over18\pi s^2_Wc^2_W} 
\log{-u\over s}\log{s\over M^2_W}
\nonumber\\
&&
-~{\alpha(1-10c^2_W)\over36\pi s^2_Wc^2_W} 
\log{-t\over s}\log{s\over M^2_W}
\eqa

\bq
c^{RG}=-~{\alpha\tilde{\beta^0}\over\pi s^2_W}\ \gaugelog
\eq

There appear important
differences between SM, Split, "m.l. MSSM", "h.s. MSSM"
in gauge, Yukawa, SUSY QCD and RG terms:\\

SM:  $n_{qq}=3$, $\eta_Y=1,$ $\eta'_Y=0$, $\eta=0$  and  
$\tilde{\beta^0}={19\over24}$ \\

Split:  $n_{qq}=3$, $\eta_Y=1,$ $\eta'_Y=0$, $\eta=0$  and 
$\tilde{\beta^0}={7\over24}$ \\ 

"m.l. MSSM":  $n_{qq}=2$, $\eta_Y=\eta'_Y=2$, $\eta=1$ and 
$\tilde{\beta^0}=~-~{1\over4}$ \\

"h.s. MSSM": $n_{qq}=3$, $\eta_Y=1$, $\eta'_Y=1$, $\eta=0$,  and 
$\tilde{\beta^0}={1\over8}$ \\

{ \bf 4) 
\underline{$u\bar d\to t\bar b$ with $W$ exchange in the $s$ channel}}

\bqa
{d\sigma\over d\cos\theta}&=&
{\pi\alpha^2
s\over32s^4_W(s-M^2_W)^2}\{~(1+\cos\theta)^2
[1+c^{ew}(b\bar b)_L+c^{ew}(u\bar u)_L+c^{ew}(d\bar d)_L
+c^{ew}(t\bar t)_L
\nonumber\\
&&+4c^{SQCD}(q\bar q)+2c^{ang}+2c^{RG}]~+~{m^2_t\over
s}\sin^2\theta~\}
\eqa

with the specific coefficients:

\bqa
c^{ang}&=&-~{\alpha\over8\pi}\ \log\frac{s}{M_W^2}
\{[4(Q_dQ_b+Q_{u}Q_{t})
+{g^Z_{dL}g^Z_{bL}+g^Z_{uL}g^Z_{tL}\over s^2_Wc^2_W}]~\log{-t\over s}]
\nonumber\\
&&-[4(Q_dQ_{t}+Q_{u}Q_{b})
+{g^Z_{dL}g^Z_{tL}+g^Z_{uL}g^Z_{bL}\over s^2_Wc^2_W}]~\log{-u\over s}]
\}
\eqa
\bq
c^{RG}=-~{\alpha\tilde{\beta^0}\over\pi s^2_W}\ \gaugelog
\eq
the $\tilde{\beta^0}$ quantity being given as above in the various models.\par
As a final process to be possibly considered in our set of "LHC 
candidates" we shall also mention the production of (light) 
chargino pairs. It should be mentioned that for this process 
the difference between Split and "no-Split" effects is reduced, 
being mostly due to variations of gauge and RG effects 
(no Yukawa components from virtual Higgs exchanges are now involved, 
therefore the difference of effects between a  superheavy Higgs and a 
"moderately light" one does not contain the $\tan^2\beta$ 
enhancement). This suggests from the beginning that accurate analyses 
of this process in the c.m. 1 TeV energy range could not be 
considered as promising candidate Split indicators. 
Since our numerical analysis of Section 3 
will confirm this (negative) expectation, we shall write the relevant 
numerical expressions for the chargino case in the Appendix B, 
thus avoiding to make Section 2A too (uselessly) long. 
Analogous considerations would apply 
for the production of neutralino pairs, 
that will not be considered in this Section 2A.\par
Our lists for the considered LHC processes are now completed. 
To make them more meaningful, a precise numerical analysis 
would now be appropriate. We shall perform it in detail in Section 3, 
where an assumed hypothetical scenario will be investigated 
starting from the expressions that we have 
written. In the following Section 2B we shall  perform a similar 
investigation of processes that might be relevant for a test of Split 
Supersymmetry at ILC.

\subsection{Pair production at ILC}

We shall now concentrate our analysis on those processes 
of pair production at the future lepton linear collider 
ILC that we consider potentially relevant 
for our investigation. To make our purposes clear, 
we shall now assume again an hypothetical situation in which, 
after the end of LHC measurements, 
only a light Higgs boson and light charginos and, possibly, 
neutralinos have been produced; 
this implies a limit on the squark masses of about 1.5-2 TeV, depending on the 
specific theoretical assumptions\cite{KM}.
For the Supersymmetric Higgses we shall assume again no evidence. 
If we want to insist on the c.m. energy region of 1 TeV as we did in 
Section 2A, this implies 
that e.g. the charged Higgs and slepton masses must be heavier than, 
roughly, 500 GeV, with possibly 
lower limits on the neutral  sectors. 
To exploit the simple Sudakov 
expansion will be then tolerable provided 
that we assume a charged Higgs and slepton masses
only slightly above 500 GeV, and this will be 
the assumption of this second 
part of Section 2, although, as we shall comment in Section 3, the possibility 
of moderately heavier Higgs bosons might also be reasonably treated. Having 
made this statement,
we shall consider only those 
processes that might exhibit "visibly" different features in the two 
(assumedly) competitor Split and "h.s. MSSM" scenarios. 
In practice, this  will limit our choice to the processes 
of production of 
heavy ($b,t$) quark pairs, that, as one can guess from our previous 
discussion, will be the only ones for which the differences 
due to variations  of the Higgs Yukawa effects will be relevant. 
For these two cases we shall 
write in the following part of the Section the relevant 
Sudakov expansions. In 
the Appendix A,C we shall write the corresponding 
expressions of three other 
processes, i.e. the production of muon,light quarks, 
chargino and neutralino 
pairs. The motivation will be mostly that of providing 
a complete list of 
Split Sudakov  logarithmic effects, even in cases for which,
as we shall show, 
the chances of identification of the model in those processes 
from an analysis 
of our kind at those energies seem to us to   be rather 
limited (this does not 
exclude the possibility of other tests at different energies, using 
measurements of a different type). Having made this statement, 
we write now 
the relevant expressions for the two heavy quark pair 
production cases.\\

In fact they are obtained from the general expression valid
for $e^+e^-\to f\bar f$ for any lepton or quark $f$:

\bqa
{d\sigma\over d\cos\theta}&=&{N_{c,f}\over32\pi s}\{
{u^2\over s^2}[|a^B_{RR}|^2(1+2c_{RR})
+|a^B_{LL}|^2(1+2c_{LL})]\nonumber\\
&&+{t^2\over s^2}[|a^B_{LR}|^2(1+2c_{LR})
+|a^B_{RL}|^2(1+2c_{RL})]\}
\eqa

\bq
a^B_{LL}={e^2\over4 s^2_Wc^2_W}[(2s^2_W-1)(2I^3_f)-2s^2_WQ_f]
~~~~~~~~~
a^B_{RR}=-~{e^2\over  c^2_W}\ Q_f
\eq
\bq
a^B_{LR}=-~{e^2\over 2 c^2_W}\ Q_f
~~~~~~~~~
a^B_{RL}=~{e^2\over  c^2_W}(I^3_f-Q_f)
\eq
\noindent
where $Q_f$, $I^3_f$ and $N_{c,f}$ are the electric charge in unit
of $|e|$, the third component of the isospin and the colour factor.\\
The one loop coefficients for each combination of chiralities
are given by the following sum:

\bq
c_{ij}=c(ee,gauge)_i+c(ff,gauge)_j+c(ff,yuk)_j
+c^{ang}_{ij}+c^{RG}_{ij}+c^{SQCD}(ff)
\eq
\noindent
with the universal parts $c(ee,gauge)_i$,
$c(ff,gauge)_j$, $c(ff,yuk)_j$, $c^{SQCD}(ff)$ 
given in Appendix A in terms of
$n(ee), n(ff)$
and $\eta$, $\eta_Y,\eta'_Y$, specific of each model,
and the following non universal parts:

\bqa
c^{ang}_{LL}&=&{\alpha\over\pi}\ \gaugelog\{
(2I^3_f){c^2_W\over s^2_W}\log{1+(2I^3_f)cos\theta\over2}
-s^2_Wc^2_W\log{t\over u}[4Q^2_f-~{2Q_fg_{eL}g_{fL}\over s^2_Wc^2_W}+
{g^2_{eL}g^2_{fL}\over 4s^4_Wc^4_W}]\}.\nonumber\\
&&.[(2s^2_W-1)(2I^3_f)-2s^2_WQ_f]^{-1}
\eqa
\bqa
c^{ang}_{RR}={\alpha c^2_W\over4\pi Q_f}\ \gaugelog\ 
\log{t\over u}\ [4Q^2_f-~{2Q_fg_{eR}g_{fR}\over s^2_Wc^2_W}+
{g^2_{eR}g^2_{fR}\over 4s^4_Wc^4_W}]
\eqa
\bqa
c^{ang}_{LR}={\alpha c^2_W\over2\pi Q_f}\ \gaugelog\
\log{t\over u}\ [4Q^2_f-~{2Q_fg_{eL}g_{fR}\over s^2_Wc^2_W}+
{g^2_{eL}g^2_{fR}\over 4s^4_Wc^4_W}]
\eqa
\bqa
c^{ang}_{RL}=-~{\alpha c^2_W\over4\pi(I^3_f-Q_f)}\ \gaugelog\
\log{t\over u}\ [4Q^2_f-~{2Q_fg_{eR}g_{fL}\over s^2_Wc^2_W}+
{g^2_{eR}g^2_{fL}\over 4s^4_Wc^4_W}]
\eqa

with
\bq
g_{eL}=2s^2_W-1~~~~~g_{eR}=2s^2_W~~~~~g_{fL}=2I^3_f-2s^2_WQ_f
~~~~~g_{fR}=-2s^2_WQ_f
\eq
and
\bqa
c^{RG}_{LL}=-\frac{1}{4\pi^2}\ \gaugelog\ [-(2I^3_f){g^4\tilde{\beta^0}\over4}
+g^{'4}\tilde{\beta^{'0}}({2I^3_f\over4}-{Q_f\over2})]
[g^{'2}({2I^3_f\over4}-{Q_f\over2})-{2I^3_f\over4}g^2]^{-1}
\eqa
\bqa
c^{RG}_{RR}=c^{RG}_{LR}=c^{RG}_{RL}
=-\frac{1}{4\pi^2}\ \gaugelog\ [g^{'2}\tilde{\beta^{'0}}]
\eqa
\noindent
using $g=e/s_W,g'=e/c_W$ and
the corresponding $\tilde{\beta^{0}},\tilde{\beta^{'0}}$
functions also given in Appendix A.\\

The specification to $f\bar f$=$t\bar t$, $b\bar b$ is obtained
with $Q_f=2/3, -1/3$, $I^3_f=1/2,-1/2$, $N_{c,f}=3$.\\

Our list of the various logarithmic Split effects that seemed to us to be 
worth being determined is now completed. In the next Section, we shall examine
two possible future situations where a comparison  of our
formulae with realistic experimental measurements might be able to 
favor one of the two competitor supersymmetric schemes that we are analyzing in this
paper.

\section{Search of specific supersymmetric signals in top production}

In this Section, we shall provide two examples of possible 
situations where a dedicated analysis of the energy 
dependence of the cross Sections of top 
production might discriminate Split from the  special competitor model
that we called ``h.s. MSSM''.
For this purpose, we shall consider two hypothetical situations, 
one at LHC and the other one at ILC, and devote the two 
Sections 3B and 3A to the discussion of the two cases.

\subsection{The LHC scenario}

As already said in the Introduction, we shall examine a 
case in which, after a certain period of time, 
say 2-3 years at reasonable luminosity, LHC had 
provided evidence of the existence of one light Higgs 
and of light charginos (and, possibly, neutralinos or gluinos), 
but no evidence of any of the 
remaining supersymmetric sparticles (other Higgs bosons, 
squarks, sleptons). 
We shall assume, following Ref.\cite{KM}, that this lack of 
discoveries can be 
translated into lower limits for the various masses of the
approximate value of $\simeq1.5$  TeV for the squarks and of 
~ 400 GeV for the 
various Higgses, although these values (in particular 
those for the squarks) could be
easily reasonably modified. Starting from this scenario, we have 
computed the energy distributions of the cross sections 
for the 4 processes listed in Section 2A, 
in an energy range around 1 TeV. In fact, a preliminary 
explanation of the 
simplifications that we have used seems now appropriate. 
First of all, we have given the formulae for the basic 
partonic components of the various 
processes. The translation of those expressions into more 
experimentally meaningful observables has been fully discussed in 
refs. \cite{quark,ttb}. In ref. \cite{ttb} the
translation from c.m. initial parton energy to final pair 
invariant mass has also been numerically analyzed, and a similar study 
for the various single top production processes is being carried on.
The formulae for the calculation of the initial proton-proton 
differential cross section are
known and have also been used in ref. \cite{quark,ttb}, using the most 
recent available distribution functions calculations.
Having made this premise, given the quite preliminary
nature of this paper, we shall be limited to the 
analysis of the c.m. energy 
dependence of the elementary partonic processes, 
and discuss its relevant, sometimes promising, 
features in the following part of this paper.

Figs. 1-4 show the percentage effects on the various top-antitop and single 
top production cross sections, at variable $\sqrt{s} =~ 1$ TeV, of Split 
Supersymmetry and of other possible alternatives. One sees immediately that,
as anticipated, the shapes of Split SUSY and of the Standard Model 
practically coincide. One also notices the big difference (systematically 
larger than 10 \%, particularly for large $\tan\beta$) between the Split and 
the "light MSSM" effects. What is relevant for our analysis is the existence 
of a sizable difference between the effects of Split and those of considered 
competitor "light Higgses MSSM" scenario. In this respect, we must provide at 
this point an explanation. In this preliminary study, we have ignored for the 
latter scenario the possible residual virtual vertex  effects of the third 
family squarks (assumedly heavier than $\sim 1.5$ TeV). In this way, we obtain a 
difference of effects that can be, for large $\tan\beta$ and $\sqrt{s} \gtrsim 1$ TeV, 
of approximately 3-4\%. This value is quite likely to be a pessimistic 
(i.e. too small) one. We expect in fact that a rigorous calculation of the 
effects of the third family squarks in the 1.5 TeV mass range  should add a 
negative contribution to the effect (thus increasing it), since we know that 
this contribution would be substantial and negative for squarks masses of 
about 400 GeV, and we do not think that it might change dramatically 
increasing the masses to the 1.5 TeV limit. To make this statement
less qualitative, a complete calculation of that contribution is requested, 
that will be the aim of a following dedicated paper. For the moment, we retain 
this preliminary result and comment it briefly. Clearly, in order to 
appreciate a difference of the five percent size, measurements of the involved 
cross sections at the same overall precision are postulated. This would 
require a dedicated work to improve the available theoretical and experimental 
accuracies, that can be estimated at the moment to be of the order of an 
overall ~twenty percent \cite{CERNYB,ttb}. This effort appears to us, generally speaking, 
quite auspicable since it would allow to reach a successful final goal of the 
measurements even if "only" SM tests were performable. In fact, as stressed in 
Ref. \cite{CERNYB}, with such an accuracy one would obtain a determination of the top 
mass to the 1 TeV precision (with remarkable benefits for various SM tests) 
and of the CKM $V_{tb}$ coupling to the five percent level. 
In this sense, we feel 
that, if extra strong motivations existed, an effort to reduce the overall 
error to the (extreme) five percent level might deserve some 
consideration.

To conclude this Section, we have computed in the following Figures 5 and 6 the 
analogous effects for production of chargino (assumed to be in the ~400 GeV 
mass region) pairs. As one can see, the effects of Split and of "h.s. MSSM" are 
in this case essentially identical. We should remark that, in our analysis, we 
have considered (a) the $\cos\theta$ integrated cross section and (b) the sum of 
the three possible chargino pairs production, thus avoiding to have to 
consider additional parameters like mixing angles (which are highly
model dependent and could hide the main features of the scenarios we
want to identify) as explained in
Appendix B. We cannot exclude that a 
consideration of these neglected possibilities may lead to less invisible effects, 
although we think that at LHC this  would be rather difficult.

\subsection{THE ILC SCENARIO}

To conclude our investigation, we have considered a case of possible ambiguity 
between the two considered scenarios that might arise at a future 
International Linear Collider (ILC) for the extreme value $\sqrt{s} =~1$ TeV. This 
requires a preliminary discussion on the assumed lower limits for the various 
involved masses. For what concerns the squarks, we shall retain the assumed 
negative LHC limits of Section 2A; for the Higgses and the sleptons, given the 
fact that we imagine to perform measurements at ~1 TeV having no evidence of 
these particles, we must assume a limit of, roughly, 500-550 GeV,
at least for the charged
sector. Since we want to retain the simple features of a Sudakov expansion, 
that requires $\sqrt{s}$ to be sufficiently larger than the involved masses, we shall
fix the minimum value 550 GeV for the latter quantities. Concerning the 
validity of the 'asymptotic" logarithmic expansion, we shall accept that it 
still retains the reliability that it had for the lower limit of 400 GeV used 
in Section 2A, and for this preliminary qualitative analysis we shall not try 
to
estimate the possible corrections coming from next-to next to leading terms of 
the expansion. Note that, differently from the situation at LHC where the 
sleptons did not contribute to the various effects independently of their 
mass, this time there will be a priori a slepton contribution of gauge origin
in the leptonic vertices. 

After these premises, we show now in the next Figures 7,8,9,10 the results of 
the comparison for the processes of production of heavy ($b,t$) quarks, of 
charged leptons and of charginos pairs. Again, we drew the curves that would 
correspond to the Standard Model and to the "moderately light" MSSM scenario 
mostly for sake of comparison. For what concerns the difference between Split 
Supersymmetry and the "heavy squarks MSSM", the following two statements appear 
to us, at this point, justified. The first one is that there appear  now two 
processes, the production of bottom-antibottom and the production of 
top-antitop pairs, shown in Figs. 7,8, where the relative difference between 
the two scenarios at $\sqrt{s} =~1$ TeV approaches a value of approximately five 
percent for large $\tan\beta$. 
Again, for the same reasons that we discussed in Section 2A, we have 
motivations to believe that this estimate might be a pessimistic one, and that 
the additional effects of the ~1.5 TeV squarks should increase the overall
difference. Assuming a final level of accuracy at ILC at the few  percent 
level, these differences should be visible. Having in mind the problems that 
bottom production had to meet at LHC as a consequence of scale QCD 
uncertainties~\cite{CERNYB}, problems that were greatly reduced in  top production 
thanks to the
lack of hadronization in the top decays, we feel that, even at ILC, top 
production would be a better candidate for the purposes of a high precision 
measurement at the few percent level. If this were the case, a choice between 
the two considered scenarios would become realistically performable.

An apparently less optimistic picture would be provided, in our approach, by a 
similar analysis performed in the case of production of charged leptons and of 
charginos pairs. This statement is supported by an inspection of the last 
corresponding Figures 9,10 that evidentiate a difference of effects between 
the two scenarios of at most two percent, i.e. at the realistic precision 
limit of the measurements. While for the production of muons this result is 
independent  of the assumptions on the squark masses and of $\tan\beta$, for the 
chargino case we believe that a number of comments would be appropriate. First 
of all, one notices an effect of possibly two percent (for large $\tan\beta$) that 
was absent in the analysis done at LHC for the same production  process. The 
technical reason is that at ILC there is a contribution, as we said, from the 
"reasonably light" slepton gauge vertices that was absent at LHC. The second 
point is that, once again, the estimated effect of chargino production is 
likely to be pessimistic, since the 1.5-2 TeV  third family squark 
contribution has not been added. Finally, as in the LHC case, we did not 
consider the possibility of measuring differential cross sections or 
separate charginos pairs production, as done in Refs. \cite{Zhu,KPRC}, at fixed 
energy. Our analysis is based in fact on the determination of the energy 
dependence of
the various quantities at the one-loop level, and tries to avoid, summing over 
the three final charginos states, the introduction of extra parameters like 
e.g. mixing angles (see Appendix B). 
We cannot exclude the possibility that an alternative 
study of chargino production, performed at the one-loop level, provides 
possible useful information for the detection of possible Split signals as 
derived at Born level in Refs. \cite{Zhu,KPRC}. An analysis of this type is already 
being carried on.

\section{Conclusions}

The aim of this preliminary paper was twofold. On one side, we wished to 
compute and to list all the relevant logarithmic expansions of Sudakov kind 
for Split Supersymmetry at energies and for processes where they could be 
measured and, therefore, tested. 
This was done in the paper, and in principle one might
consider the shape of the energy dependence of the chosen processes as a 
"necessary condition " to be met by Split Supersymmetry, in the sense that the 
appearance of a clearly different energy dependence would be a rather strong 
evidence against the scenario. In case of positive experimental evidence from 
the energy distribution, since other
supersymmetric scenarios might exhibit a similar energy dependences, at a 
second stage, we wished to show that, 
within the set of allowed production processes, there exists a subset whose 
accurate measurements might discriminate Split Supersymmetry from possible 
competitor ``TeV" supersymmetry scenarios. Our conclusion is that there appears 
indeed to exist a special subset of processes, 
those of production of top-antitop pairs and of single top production at LHC 
and those of top pairs and, possibly, chargino pairs (with a minor chance, in 
our opinion, for
charged lepton pairs) at ILC, that would allow, under certain reasonable 
circumstances, to discriminate Split from another competitor supersymmetric 
scenario. 
Given 
the fact that top-antitop production and single top production at LHC are 
already considered as extremely interesting processes, we believe that our 
analysis, simply, adds another amount of interest to their measurements, thus 
stimulating a continuation of the dedicated theoretical and experimental 
studies that already exist\cite{CERNYB}. 
For the ILC processes, we feel that, if
this turned out to be the case at the time of the future machine performances, 
the possibility of reaching the extreme (at the "below two percent" level) 
accuracy requested to support the revolutionary Split proposal by 
investigation of the processes that we listed might be seriously taken into 
consideration. 
From our side, the generalization 
of our preliminary results to a less specific supersymmetric alternative 
scenario and the rigorous calculation of some numerical details that were not 
examined in this paper, as we already stressed, is already being carried on.

\appendix

\section{Logarithmic coefficients in SM, MSSM and Split}

We follow the classification of logarithmic terms made in
refs.\cite{sud,quark} 
and for each type of term we indicate the
modification arising when passing from the MSSM case with a light SUSY
scale to the Split case with heavy scalars.\\

{\bf \underline{1)  Electroweak RG corrections}}\\

These terms arise from gauge boson or gaugino self-energies.
In the Split case, the bubbles containing heavy scalars
are suppressed, leading to a modification of the
$\tilde{\beta}$ functions as indicated below,
see \cite{GR}. The correction to the
Born amplitude is obtained as

\bq
A^{RG}=-~{1\over4\pi^2}[g^4\tilde{\beta^0}
{dA^{Born}\over dg^2}+g^{'4}\tilde{\beta^{'0}}
{dA^{Born}\over dg^{'2}}]\ \gaugelog
\label{crg}\eq

with\\

$\tilde{\beta^0}={19\over24},~~{7\over24},~-~{1\over4},~~{1\over8}$
~~~~~~$\tilde{\beta^{'0}}
=-~{41\over24},-~{15\over8},~-~{11\over4},-~{55\over24}$\\

in 
SM, Split, "m.l. MSSM", "h.s. MSSM", respectively.\\

{\bf \underline{2)  Universal electroweak corrections}}\\

These corrections arise from collinear and soft singularities
of one loop electroweak effects; they are usually called of Sudakov
origin and they factorize the Born amplitude in a process
independent way, depending only on the quantum numbers of
the external lines. They are written (possibly in a matrix form
when the external particles are mixed states) as:

\bq
A^{univ}=c^{ew}~A^{Born}
\eq

\vspace{1cm}

\underline{2a) External leptons and quarks pairs}\\

For a given chirality $a=L$ or $R$, the universal 
coefficients $c^{ew}$ are usually written as a sum of 
a gauge term and of a Yukawa term:

\bq
c^{ew}(f\bar f)_a=
c(f\bar f, ~gauge)_a~+~c(f\bar f,~yuk)_a
\eq

\underline{The gauge term}
arises from SM diagrams containing an internal gauge boson
associated to a fermion ($Wf,\gamma f,Zf$), 
and from SUSY diagrams containing
an internal chargino or neutralino (gaugino component)
associated to the corresponding
sfermion ($\chi^{\pm} \tilde{f},\chi^{0}  \tilde{}f$). In Split,
this second contribution is suppressed.

\bq
c(f\bar f, ~gauge)_a=b(f\bar f)_a~(n ~\log{s\over M^2_W}
-\log^2{s\over M^2_W})
\eq
\bq
b(f\bar f)_a={\alpha\over4\pi}~[~{I_f(I_f+1)\over s^2_W}
~+~{Y^2_f\over4c^2_W}~]
\eq
$I_f$ is the full isospin and $Y_f=2(Q_f-I^3_f)$ the hypercharge,
of the external fermion $f$  of chirality $a$. The single log index
for quarks is  $n=3$ in SM, h.s.MSSM and Split, 
and $n=2$ in m.l.MSSM, whereas for leptons it is 
$n=3$ in SM and Split, 
and $n=2$ in m.l.MSSM and h.s.MSSM.\\

\underline{The Yukawa term} arises from (SM and SUSY) diagrams with
internal Goldstones and Higgs bosons associated to a
fermions($G^{\pm}f,G^0f$),
($H^{\pm}f,h^0f,H^0f,A^0f$) and from SUSY diagrams with 
chargino or neutralino (Higgsino components)
 associated to the corresponding sfermion ($\chi^{\pm}
\tilde f,\chi^{0} \tilde f$). In Split,
the contributions with heavy $H^{\pm},H^0,A^0$
and heavy $\tilde f$  are suppressed. In all cases
this Yukawa contribution is non negligible
only for external $t,b$ quarks.

\bqa
c(b\bar b, ~yuk)_L=
c(t\bar t, ~yuk)_L&=&-~{\alpha\over16\pi s^2_W}~
\ \log{s\over M^2_W}\ 
[{m^2_t\over M^2_W}(\eta_Y+\eta'_Y\cot^2\beta)\nonumber\\
&&
+{m^2_b\over M^2_W}(\eta_Y+\eta'_Y \tan^2\beta)]
\eqa
\bq
c(b\bar b, ~yuk)_R=
-~{\alpha \over8\pi s^2_W}\ \log{s\over M^2_W}\
[{m^2_b\over M^2_W}(\eta_Y+\eta'_Y \tan^2\beta)]
\eq
\bq
c(t\bar t, ~yuk)_R=
-~{\alpha \over8\pi s^2_W}~\ \log{s\over M^2_W}\
[{m^2_t\over M^2_W}(\eta_Y+\eta'_Y \cot^2\beta)]
\eq
\noindent
in both SM and Split: $\eta_Y=1$, $\eta'_Y=0$,\\
in "m.l. MSSM": $\eta_Y=\eta'_Y=2$,\\ whereas
"h.s. MSSM": $\eta_Y=1$, $\eta'_Y=1$.\\

\underline{2b) External  transverse $W^{\pm}_T,~\gamma,~Z_T$ }\\

The contributions are in principle due to all possible pairs 
of internal particles (fermions,
sfermions, gauge bosons, gauginos, Higgses and Higgsinos),
however, see Ref.\cite{sud,DP}, 
because of a typical gauge cancellation
between splitting and parameter renormalization, the contributions
to the single log cancels, and it remains only the quadratic
log contribution feeded by the three gauge boson coupling, which is a
pure SM term. So, owing to this cancellation,
the universal coefficient is the same in all cases (SM, Split,
"m.l. MSSM" or "h.s. MSSM"):

\begin{equation}
c(W)=~{\alpha\over 4\pi s^2_W}[-\log^2{s\over M^2_W}]~~~~~c(B)=~0
\end{equation}

The term $c(W)$ directly applies to the charged $W^+$ case. In the
case of neutral photon and $Z$ external lines one has to take care of
the $W,B$ mixing and one can write the matrix rule:

\begin{equation}
A^{univ}_i = \sum_{j}~ \ c_{ij}\ A^{Born}_j
\end{equation}
\noindent
where $i,j$ refers to $\gamma$ or $Z$ with the coefficients

\begin{equation}
c_{\gamma\gamma}=~{1\over4}[-\log^2{s\over M^2_W}]~~~~~~~~~
c_{ZZ}=~{c^2_W\over4s^2_W}[-\log^2{s\over M^2_W}]
~~~~~~~~~ c_{\gamma Z}=~{c_W\over4s_W}[-\log^2{s\over M^2_W}]
\end{equation}

\underline{2c) External charged and 
neutral Higgs and Goldstone bosons}\\

This concerns now the set of charged and neutral Higgs bosons
$S\equiv H^{\pm},h^0,H^0,A^0$
and of charged and neutral Goldstone states $G^{\pm},G^0$, 
these latter ones being, at high energy, equivalent 
to the longitudinal (helicity $0$) $W^{\pm}_0$ and $Z_0$
components.\par
In this case also the coefficients result from a sum of universal
"splitting" terms (squared logs and single logs of gauge and
Yukawa type) and, depending of the process, 
of Parameter Renormalization terms, arising from Born Yukawa
couplings, which only contribute to the single log part.\par
In this paper we only need to consider the process
$gb\to tG^-$ which, at high energy, is equivalent to the
longitudinal (helicity zero) component of $gb\to tW^-$.
The Parameter Renormalization contribution \cite{DP}
to the $btG^-$ coupling (proportional to $m_t$) appearing
at Born level, cancels the "universal "splitting" single log,
much like in the transverse gauge boson case.
This cancellation is complete in the m.l. MSSM, but is incomplete
in the other models considered here (SM, Split, h.s. MSSM)
in which the squarks are too heavy. The net result can be written
as:

\bqa
c^{ew}(G^-_0)
&=&{\alpha\over\pi} {1+2c^2_W\over32s^2_Wc^2_W}
[n_{G}~\gaugelog-\gaugeloglog]
\eqa
\noindent
where, in SM, Split and h.s. MSSM  $n_{G}=4$, and
in "m.l. MSSM" $n_{G}=0$.\\

\underline{2d) External charginos and neutralinos}\\

It is convenient to treat separately the gaugino components
(specified by the mixing
elements $Z^{\pm}_{1i}$, $Z^{N}_{1i}$, or $Z^{N}_{2i}$)
and the Higssino components 
(specified by the mixing elements $Z^{\pm}_{2i}$,
$Z^{N}_{3i}$, or $Z^{N}_{4i}$)
according to the notations of \cite{Rosiek}. We will compare the
universal coefficients in the Split, "m.l.MSSM" and 
"h.s. MSSM" cases.\par

The gaugino components get
corrections totally similar to the ones of transverse gauge bosons,
i.e. pure quadratic logs. This feature remains valid in all models, 
because as explained above, it is of pure gauge nature 
(3  gauge boson couplings
and its SUSY gauge boson-gaugino-gaugino counterpart).
However the higgsino components get both gauge and
Yukawa terms. The gauge terms arise from internal 
(gauge boson,Higgsino)
and from internal (gaugino,Higgs or Goldstone) contributions. 
The Split case differs from the "m.l.MSSM" and 
"h.s. MSSM" cases by the absence of the superheavy Higgs
part in this second type of gauge terms (gaugino,Higgs or Goldstone).
This leads to a peculiar modification of the single log
index which is $n=2$ in MSSM. It becomes 
$n=3-\cos^2\beta$ 
for R-charginos ($Z^{-}_{2i}$ component)
and for $Z^{N}_{3i}$ Higgsino component of neutralinos,
and  $n=3-\sin^2\beta$ for L-charginos ($Z^{+}_{2i}$ Higgsino
component) and for
$Z^{N}_{4i}$ Higgsino component of
neutralinos; the contributions
which are now suppressed in Split were those giving,
in the MSSM, the complementary
$\sin^2\beta$ and $\cos^2\beta$ parts leading to $n=2$.
The Yukawa terms arising from ($f\tilde{f}$) contributions
are totally suppressed in the Split case because of the superheavy
sfermions.\par
The result can be written in matrix form 
for an external chargino line:

\begin{eqnarray}
&&c(\chi^{+}_i\chi^{+}_j)=
{1\over 4s^2_W}~[-\log^2{s\over M^2_W}]~(Z^{+}_{1i}Z^{+*}_{1j}P_L
+Z^{-*}_{1i}Z^{-}_{1j}P_R)+\nonumber\\
&& 
{(1+2c^2_W)\over 32s^2_Wc^2_W}~(
[~n_{\chi\chi L}\log{s\over M^2_W}-\log^2{s\over M^2_W}~]
~Z^{+}_{2i}Z^{+*}_{2j}P_L
+
[~n_{\chi\chi R}\log{s\over M^2_W}-\log^2{s\over M^2_W}~]
~Z^{-*}_{2i}Z^{-}_{2j}P_R)
\nonumber\\
&&
-\eta~{3\over 16s^2_W M^2_W}~[m^2_t(1+\cot^2\beta)
~Z^{+}_{2i}Z^{+*}_{2j}P_L+m^2_b(1+\tan^2\beta)
~Z^{-*}_{2i}Z^{-}_{2j}P_R]\ \log{s\over M^2}
\end{eqnarray}
\noindent
and for an external neutralino line

\begin{eqnarray}
&&c_{\chi^{0}_i \chi^{0}_j}=
{1\over 4s^2_W}~[-\log^2{s\over M^2_W}]~(Z^{N*}_{2i}Z^{N}_{2j}P_L+
Z^{N}_{2i}Z^{N*}_{2j}P_R)
\nonumber\\
&& 
+{(1+2c^2_W)\over 32s^2_Wc^2_W}]
(~[~n_{44}\log{s\over M^2_W}-\log^2{s\over M^2_W}~]
~(Z^{N*}_{4i}Z^{N}_{4j}P_L+Z^{N}_{4i}Z^{N*}_{4j}P_R)\nonumber\\
&&
+
[~n_{33}\log{s\over M^2_W}-\log^2{s\over M^2_W}~]
~(Z^{N*}_{3i}Z^{N}_{3j}P_L+Z^{N}_{3i}Z^{N*}_{3j}P_R))\nonumber\\
&&
-\eta~({3\over 16s^2_W})[m^2_t(1+\cot^2\beta)~
(Z^{N*}_{4i}Z^{N}_{4j}P_L+Z^{N}_{4i}Z^{N*}_{4j}P_R)\nonumber\\
&&
+m^2_b(1+\tan^2\beta)~(Z^{N*}_{3i}Z^{N}_{3j}P_L
+Z^{N}_{3i}Z^{N*}_{3j}P_R)]\ \log{s\over M^2}
\label{neut}\end{eqnarray}
\noindent
with
\bq
n_{\chi\chi R}=n_{33}=2~~{\rm ("m.l. MSSM", "h.s. MSSM")}
~~~{\rm or}~~~
3-\cos^2\beta~~{\rm (Split)}
\eq
\bq
n_{\chi\chi L}=n_{44}=2~~{\rm ("m.l. MSSM", "h.s. MSSM")}~~~
{\rm or}~~~3-\sin^2\beta~~{\rm (Split)}
\eq
\bq
\eta=1~~{\rm ("m.l. MSSM")}~~~{\rm or}~~~0~~{\rm
(Split,"h.s. MSSM")}
\eq

{\bf 3) Angular dependent terms}\\

These additional peculiar terms
are just residual parts of squared log contribution
$\log^2|x|$ arising from the soft-collinear singularity 
of diagrams involving gauge boson
exchanges when
$x\equiv t\simeq-~{s\over2}(1-\cos\theta)$ or 
$u\simeq-~{s\over2}(1+\cos\theta)$ (the crossed channel Mandelstam
parameters).
After having extracted from $\log^2|x|$
the universal angular independent part $\log^2s$,
there remains an angular-dependent, process-dependent
single log term of the type

$$2\log{|x|\over s}\ \gaugelog.$$ 

There are only few such terms (typical triangle and box diagrams
with gauge boson exchanges), which are all of pure
SM gauge origin and have been explicitly computed for all considered
processes\cite{quark,sud}.\\

{\bf 4) SUSY-QCD terms}\\

In the SM case, when one considers
processes with external quarks and gluons one has to take into
account specific QCD corrections (virtual effects
and gluon bremsstrahlung effects) which are not discussed
in this paper. However, in the MSSM, at one loop logarithmic
order, the additional SUSY contributions that appear
are easily identified.\par 
For an external quark line they arise from (gluino,squark) 
contributions and for an external squark line
from (gluino,quark) contributions\cite{quark}. 
For external gluon or gluino
lines the single log term cancels as already noticed for electroweak
gauge bosons, and the quadratic log is of pure standard QCD origin and
is combined with soft gluon emission effects.\par
In the Split case, with superheavy squarks, one has just to
consider the case of an external ($q\bar q$) pair, but the 
SUSY (gluino,squark) is now suppressed. So one can write
in general

\bq
c^{SQCD}(q\bar q)=-\eta~{\alpha_s\over3\pi}\log{s\over M^2}
\label{csqcd}\eq
\noindent
with  $\eta=1$ in "m.l. MSSM" and $\eta=0$ in Split
and "h.s. MSSM" .

\section{Chargino pair production}

Using the coefficients described in Section 2
and in Appendix A, we can explicitly write the result for chargino pair
production as follows. In this paper we shall only consider
the cross section for the sum of the four processes,
$f\bar f\to\chi^+_i\chi^-_j$ with $i=1,2$ and $j=1,2$.
It is easy to check that, at high energy, when mass effects of the
order $m^2/s$ are neglected, by making this summation and using
the unitarity properties of the
$Z^{\pm}_{ij}$ mixing matrices one gets rid of the highly model dependent 
mixing matrices elements. 

\bq
\sum_{ij}{d\sigma\over d\cos\theta}(f\bar f\to\chi^+_i\chi^-_j)=
{d\sigma^{\rm Higgsino}\over d\cos\theta}+
{d\sigma^{\rm Wino}\over d\cos\theta}
\eq

In order to write compact expressions, 
we define the $G(n) = n\gaugelog-\gaugeloglog$.
The \underline{Higgsino part} reads:
\bqa
{d\sigma^{\rm Higgsino}\over d\cos\theta}&=&
{\pi\alpha^2\over2sN_c}\{{u^2(2I^3_f(1-2s^2_W)+2s^2_WQ_f)^2
\over16s^4_Wc^4_Ws^2}
[1+2b(ff)_L G(n_{ff})+2b^{\rm Higgsino} G(n_{\chi\chi L})
\nonumber\\
&&+(2b^{\rm Higgsino~Yuk}_L +2b^{\rm ang,Higgsino}_{LL}  
+2b^{RG}_{LL} +2b^{SQCD}(ff))\gaugelog ]
\nonumber\\
&&+{t^2Q^2_f\over4c^4_Ws^2}
[1+2b(ff)_R G(n_{ff})+2b^{\rm Higgsino} G(n_{\chi\chi L})
\nonumber\\
&&+(2b^{\rm Higgsino~Yuk}_L  +2b^{\rm ang,Higgsino}_{RL}
+2b^{RG}_{RL} +2b^{SQCD}(ff))\gaugelog ]
\nonumber\\
&&+{t^2(2I^3_f(1-2s^2_W)+2s^2_WQ_f)^2\over16s^4_Wc^4_Ws^2}
[1+2b(ff)_L G(n_{ff})+2b^{\rm Higgsino} G(n_{\chi\chi R})
\nonumber\\
&&+(2b^{\rm Higgsino~Yuk}_R +2b^{\rm ang,Higgsino}_{LR} 
+2b^{RG}_{LR} +2b^{SQCD}(ff))\gaugelog ]
\nonumber\\
&&+{u^2Q^2_f\over4c^4_Ws^2}
[1+2b(ff)_R G(n_{ff})+2b^{\rm Higgsino} G(n_{\chi\chi R})
\nonumber\\
&&+(2b^{\rm Higgsino~Yuk}_R +2b^{\rm ang,Higgsino}_{RR}
+2b^{RG}_{RR}+2b^{SQCD}(ff))\gaugelog ]\}
\eqa
with the 1 loop coefficients in Split, m.l. MSSM and h.s. MSSM:\\

$b(ff)_{L,R}$  and $b^{SQCD}(ff)$ for initial light leptons or quarks,
with
\bq
n_{ff}=2~({\rm m.l. MSSM})~~~~{\rm or}~~~~3~({\rm Split,h.s. MSSM})
\eq

\bq
b^{\rm Higgsino}={\alpha(1+2c^2_W)\over16\pi s^2_Wc^2_W}
\eq
with
\bq
n_{\chi\chi L}=2~({\rm m.l. MSSM,h.s. MSSM})~~~~{\rm or}
~~~~3-\sin^2\beta~({\rm Split})
\eq
\bq
n_{\chi\chi R}=2~({\rm m.l. MSSM,h.s. MSSM})~~~~{\rm or}
~~~~3-\cos^2\beta~({\rm Split})
\eq
\bq
b^{\rm Higgsino~Yuk}_L=
-\eta~{3\alpha m^2_t(1+\cot^2\beta)\over8\pi s^2_WM^2_W}
~~~~~~
b^{\rm Higgsino~Yuk}_R=
-\eta~{3\alpha m^2_b(1+\tan^2\beta)\over8\pi s^2_WM^2_W}
\eq
\bq
~\eta=1~~({\rm m.l. MSSM})~~~~{\rm or}~~~\eta=0~({\rm Split,h.s. MSSM})
\eq

\bq
b^{ang,Higgsino}_{LL}=b^{ang,Hig}_{LR}={(2I^3_f)(1-2s^2_W(1-|Q_f|))
\alpha\over4\pi s^2_Wc^2_W}
[\log{u\over t}]-y_f{\alpha c^2_W\over\pi s^2_W}[\log{x_f\over s}]
\eq
with 
\bq
y_f=~{1\over1-2s^2_W(1-|Q_f|)}~~~~~~~~
x_f=-u\delta_{I^3_f,-}-t\delta_{I^3_f,+}
\eq

\bq
b^{ang,Higgsino}_{RL}=b^{ang,Hig}_{RR}={Q_f\alpha\over2\pi c^2_W}
[\log{u\over t}]
\eq

\bq
b^{RG}_{LL}=b^{RG}_{LR}=-~{\alpha s^2_W c^2_W\over
\pi(2I^3_f(1-2s^2_W)+2s^2_WQ_f)}
({(2I^3_f)\tilde\beta_0\over s^4_W}
+{(2Q_f-2I^3_f)\tilde\beta'_0\over c^4_W})
\eq
\bq
b^{RG}_{RL}=b^{RG}_{RR}=-~{\alpha \tilde\beta'_0\over\pi c^2_W}
\eq
\noindent
with the corresponding values of $\tilde\beta^0,\tilde\beta'^0$
for the various models.\\

For the \underline{Wino part} we have:
\bqa
{d\sigma^{\rm Wino}\over d\cos\theta}&=&{\pi\alpha^2 x_f\over8s^4_WsN_c}
\{1+2b(ff)_L G(n_{ff})+2b^{\rm Wino}[-\gaugeloglog]\nonumber\\
&&
+2b^{\rm ang,Wino}\gaugelog+2b^{SQCD}(ff)\gaugelog\}
\eqa
\noindent
with, from the Born part
\bq
x_f=u^2({1\over s}+{\eta\over u}\delta_{I^3_f,-})^2
+t^2({1\over s}+{\eta\over t}\delta_{I^3_f,+})^2
\eq
\noindent
always with  (the sfermion exchange being absent in Split)
\bq
\eta=1~~({\rm m.l. MSSM})~~~~{\rm or}~~~\eta=0~({\rm Split,h.s. MSSM})
\eq
\noindent
and the 1 loop coefficients
\bq
b^{\rm Wino}={\alpha\over2\pi s^2_W}
\eq

\bq
b^{ang,Wino}={(2I^3_f)\alpha\over2\pi s^2_W}
[\log{u\over t}]
-~{\alpha\over2\pi s^2_W}(2+
\eta{s\over t\delta_{I^3_f,-}+u\delta_{I^3_f,+}})[\log{-u\delta_{I^3_f,-}
-t\delta_{I^3_f,+}\over s}]~~.
\eq

\newpage

\begin{figure}
\centering
\epsfig{file=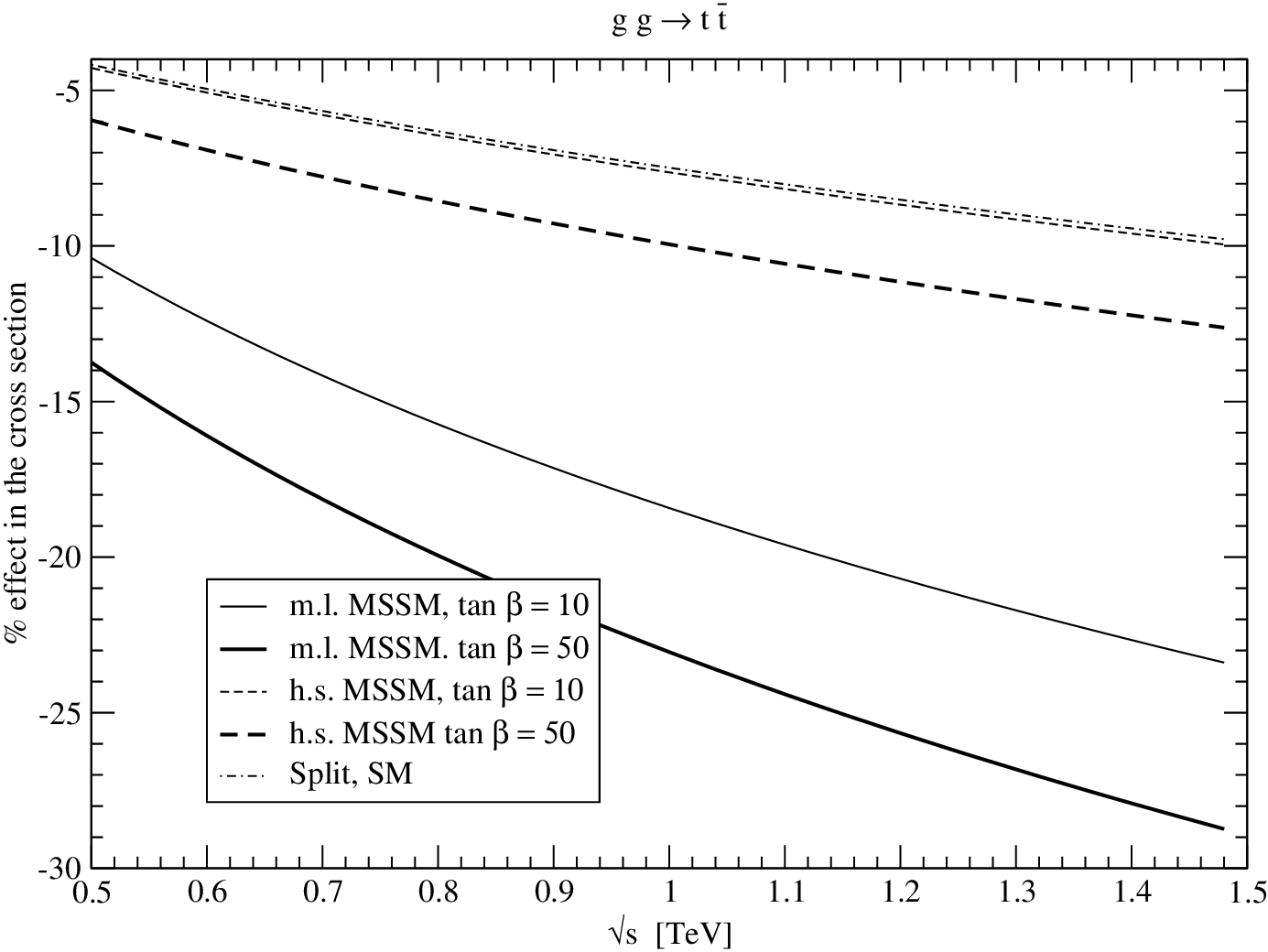,width=16cm,angle=90}
\vspace{1.5cm}
\caption{
Relative percentual effect in the cross section at partonic level 
for the process $gg\to t\overline{t}$
in various scenarios. 
Notice that there is no difference between SM and Split in our treatment.
}
\label{fig1}
\end{figure}

\newpage

\begin{figure}
\centering
\epsfig{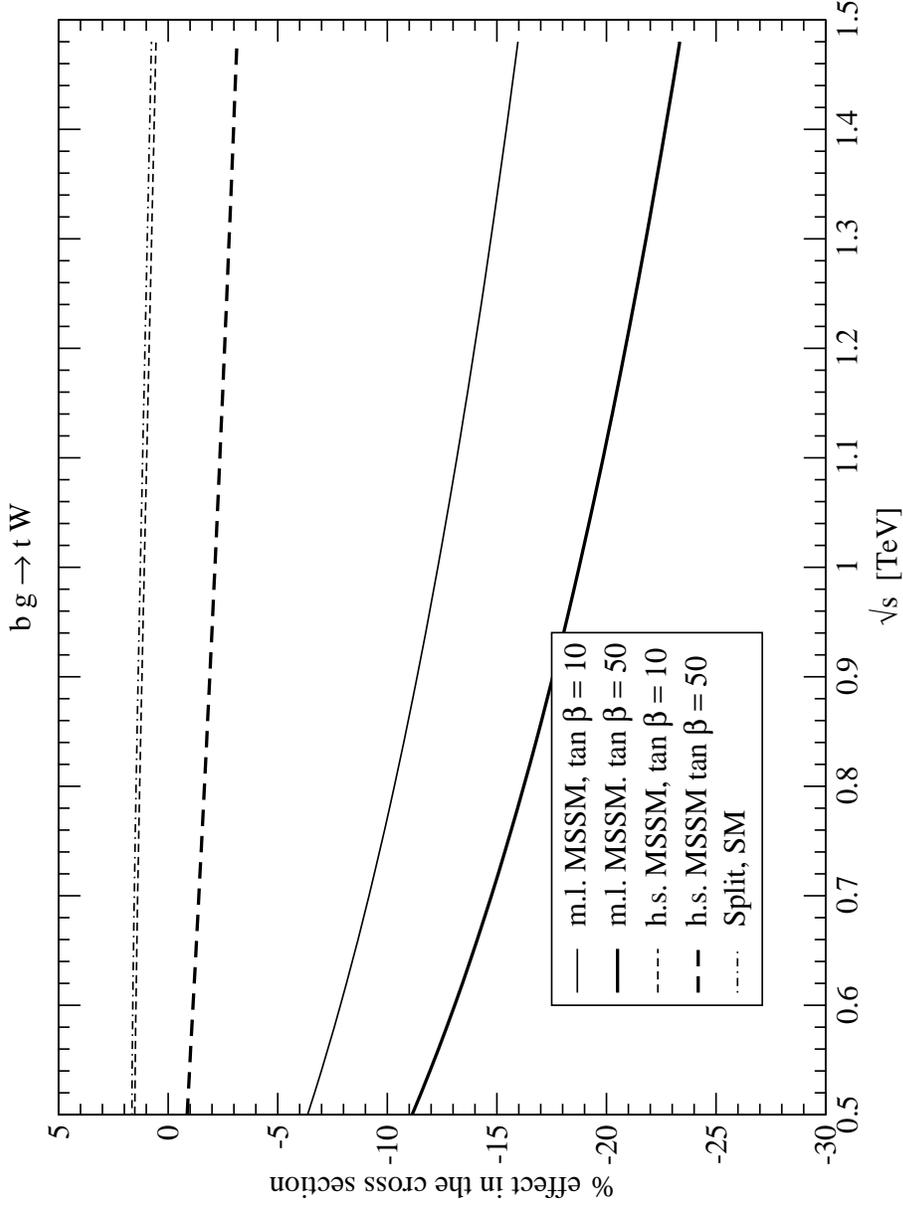}
\vspace{1.5cm}
\caption{
Relative percentual effect in the cross section at partonic level 
for the process $bg\to tW^-$
in various scenarios.
Notice that there is no difference between SM and Split in our treatment.
}
\label{fig2}
\end{figure}

\newpage

\begin{figure}
\centering
\epsfig{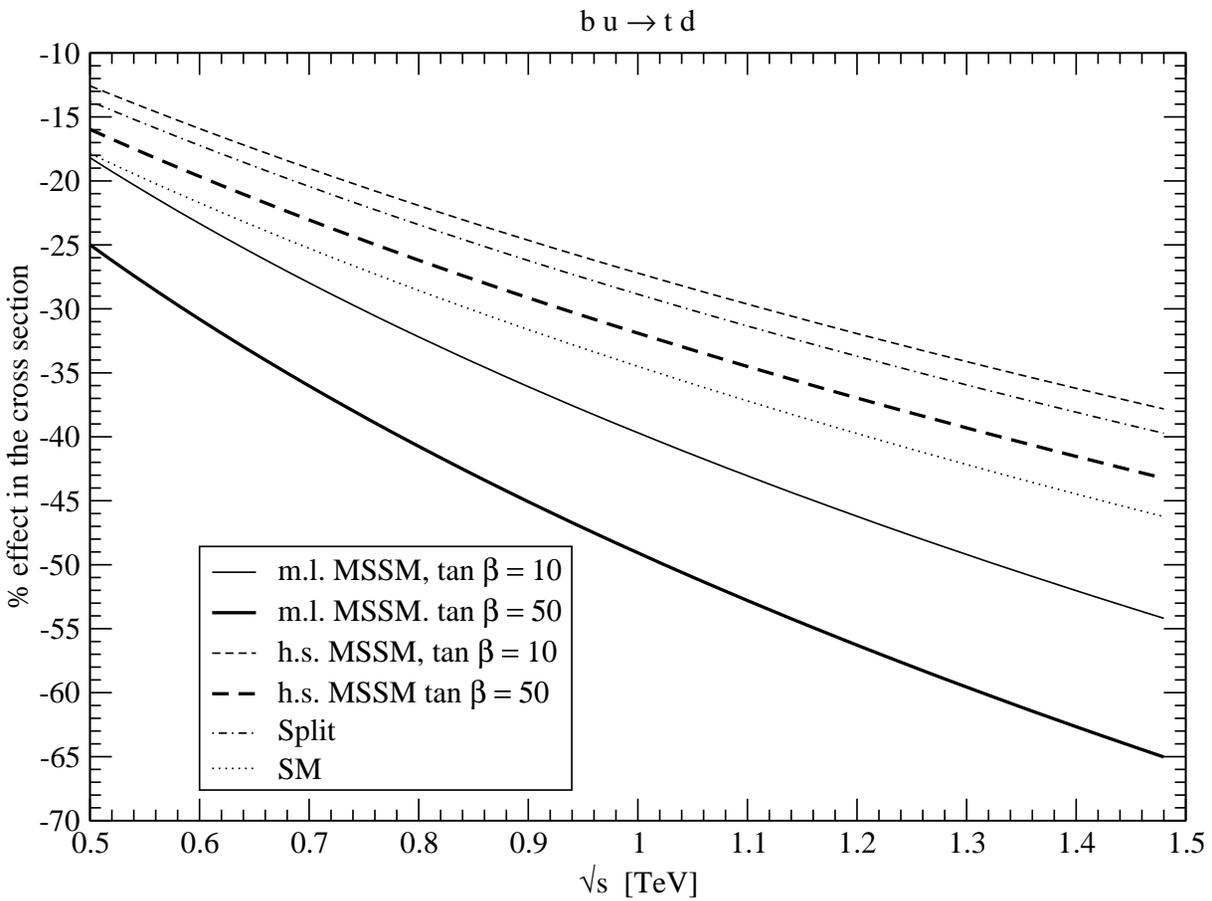}
\vspace{1.5cm}
\caption{
Relative percentual effect in the cross section at partonic level 
for the process $bu\to td$
in various scenarios.
}
\label{fig3}
\end{figure}

\newpage

\begin{figure}
\centering
\epsfig{file=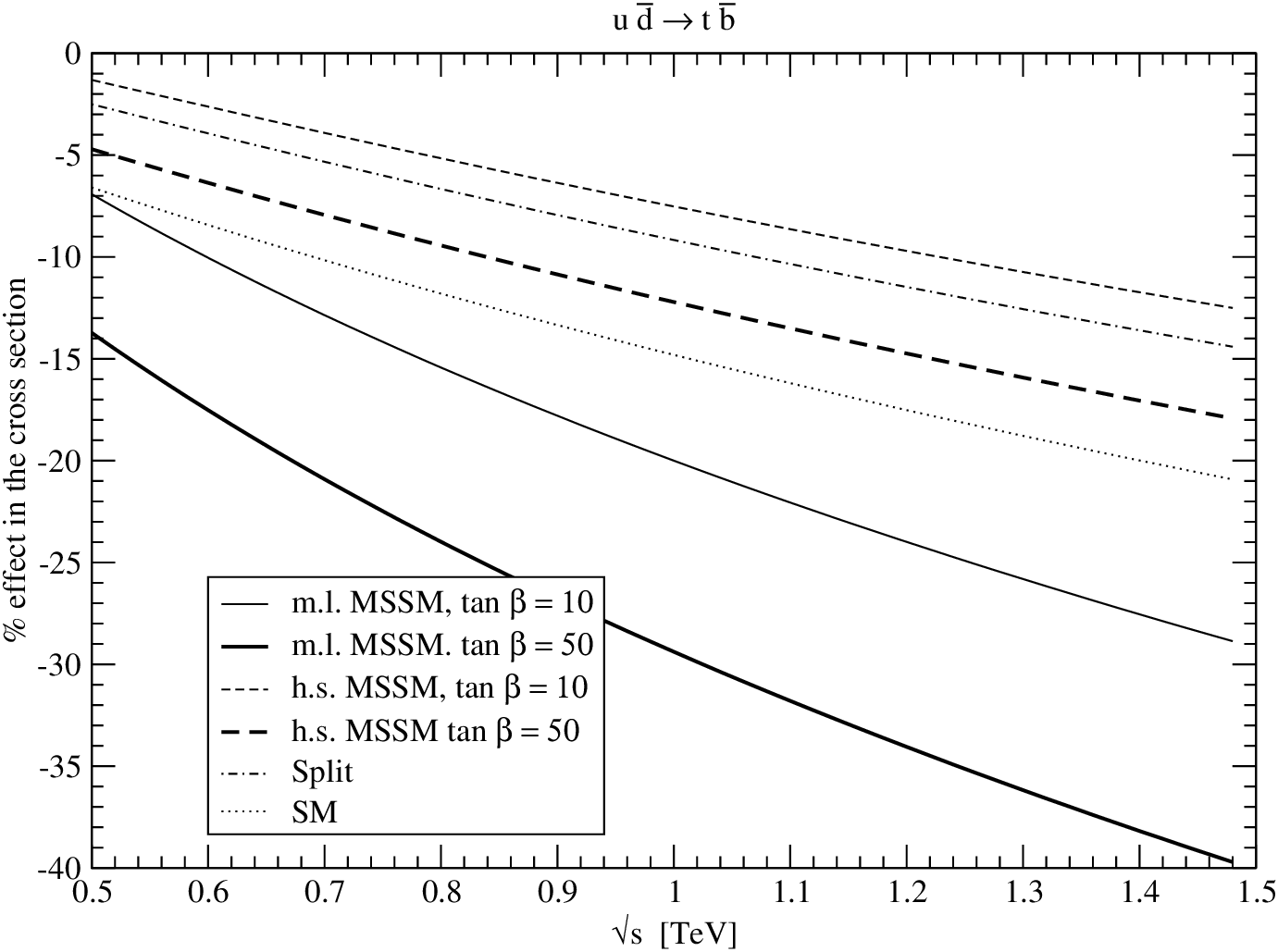,width=16cm,angle=90}
\vspace{1.5cm}
\caption{
Relative percentual effect in the cross section at partonic level 
for the process $u\overline{d}\to t\overline{b}$
in various scenarios.
}
\label{fig4}
\end{figure}

\newpage

\begin{figure}
\centering
\epsfig{file=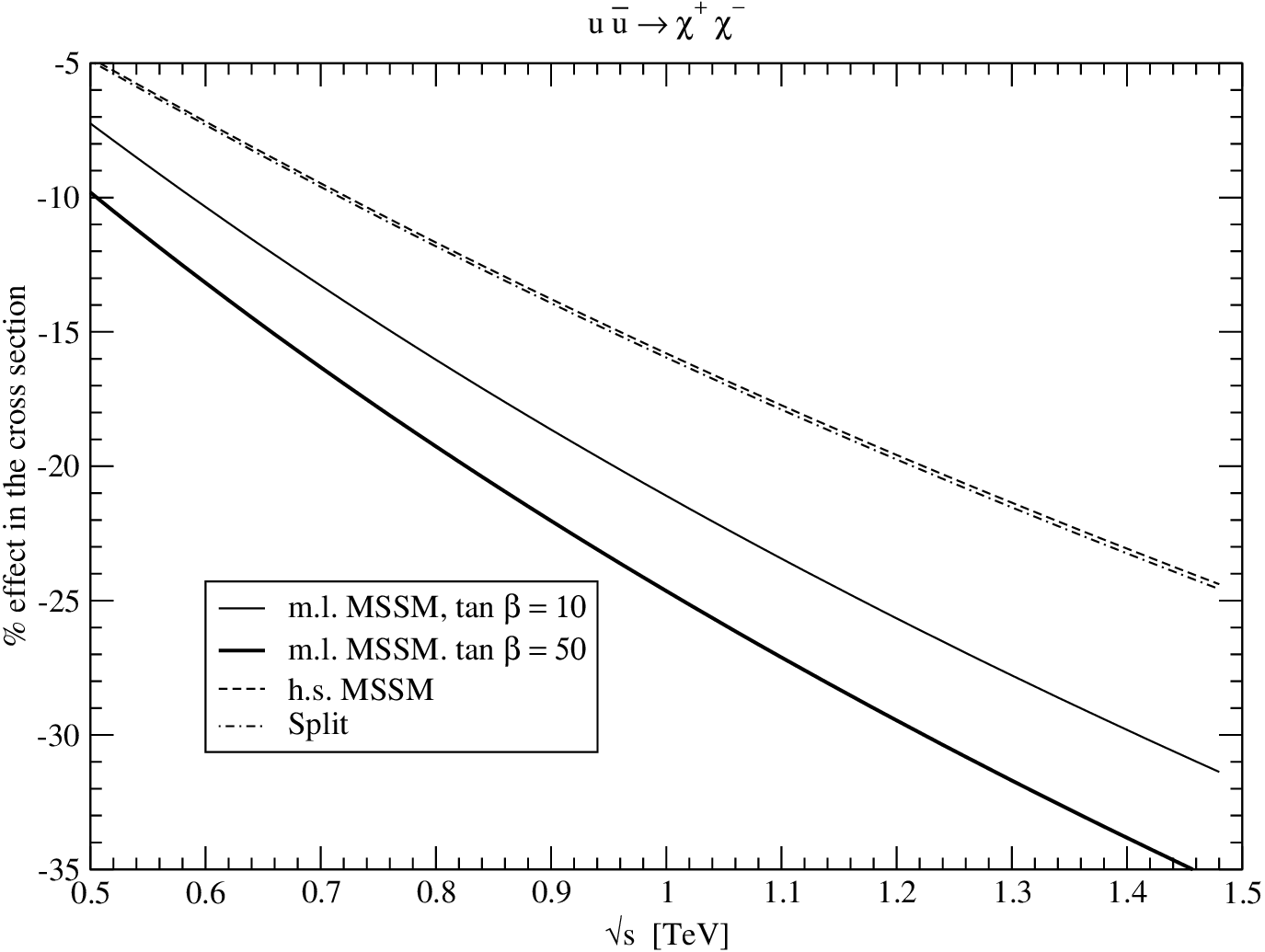,width=16cm,angle=90}
\vspace{1.5cm}
\caption{
Relative percentual effect in the cross section at partonic level 
for the process $u\overline{u}\to \chi^+\chi^-$
in various scenarios.
Notice that there is no $\tan\beta$ dependence in the model ``h.s. MSSM''.
}
\label{fig5}
\end{figure}

\newpage

\begin{figure}
\centering
\epsfig{file=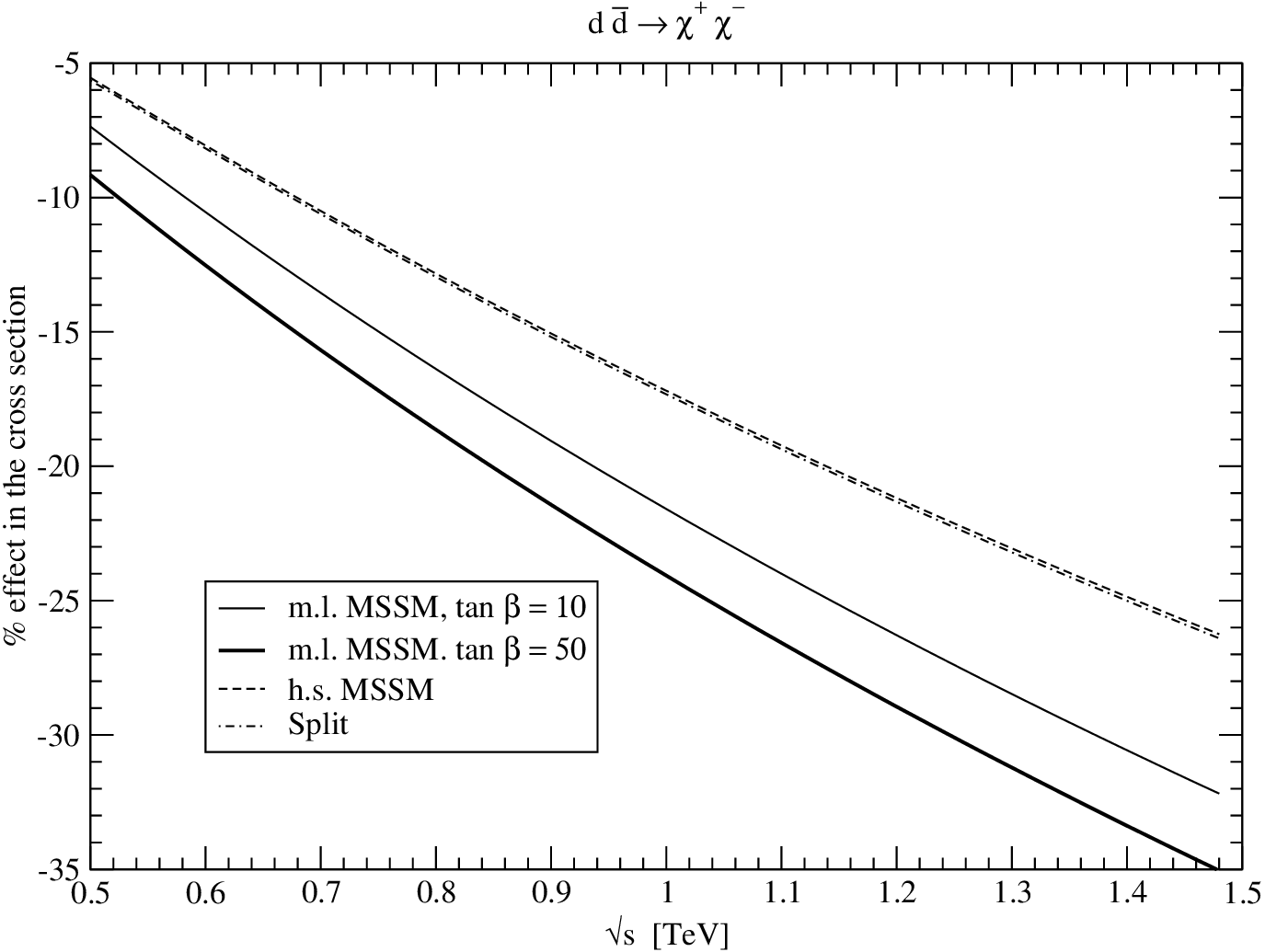,width=16cm,angle=90}
\vspace{1.5cm}
\caption{
Relative percentual effect in the cross section at partonic level 
for the process $d\overline{d}\to \chi^+\chi^-$
in various scenarios.
Notice that there is no $\tan\beta$ dependence in the model ``h.s. MSSM''.
}
\label{fig6}
\end{figure}

\newpage

\begin{figure}
\centering
\epsfig{file=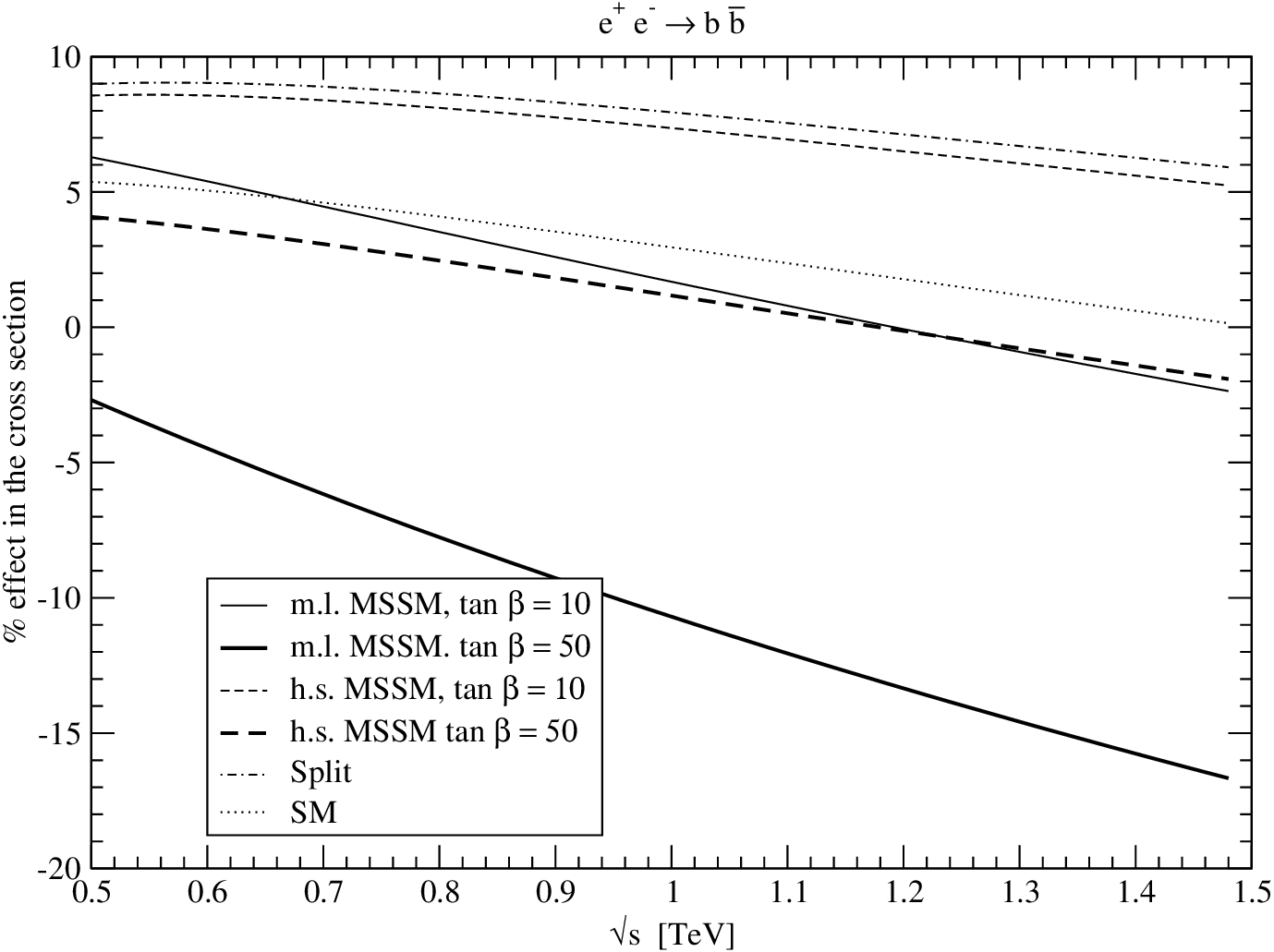,width=16cm,angle=90}
\vspace{1.5cm}
\caption{
Relative percentual effect in the cross section at partonic level 
for the process $e^+e^-\to b\overline{b}$
in various scenarios.
}
\label{fig7}
\end{figure}

\newpage

\begin{figure}
\centering
\epsfig{file=eett.eps,width=16cm,angle=90}
\vspace{1.5cm}
\caption{
Relative percentual effect in the cross section at partonic level 
for the process $e^+e^-\to t\overline{t}$
in various scenarios.
}
\label{fig8}
\end{figure}

\newpage

\begin{figure}
\centering
\epsfig{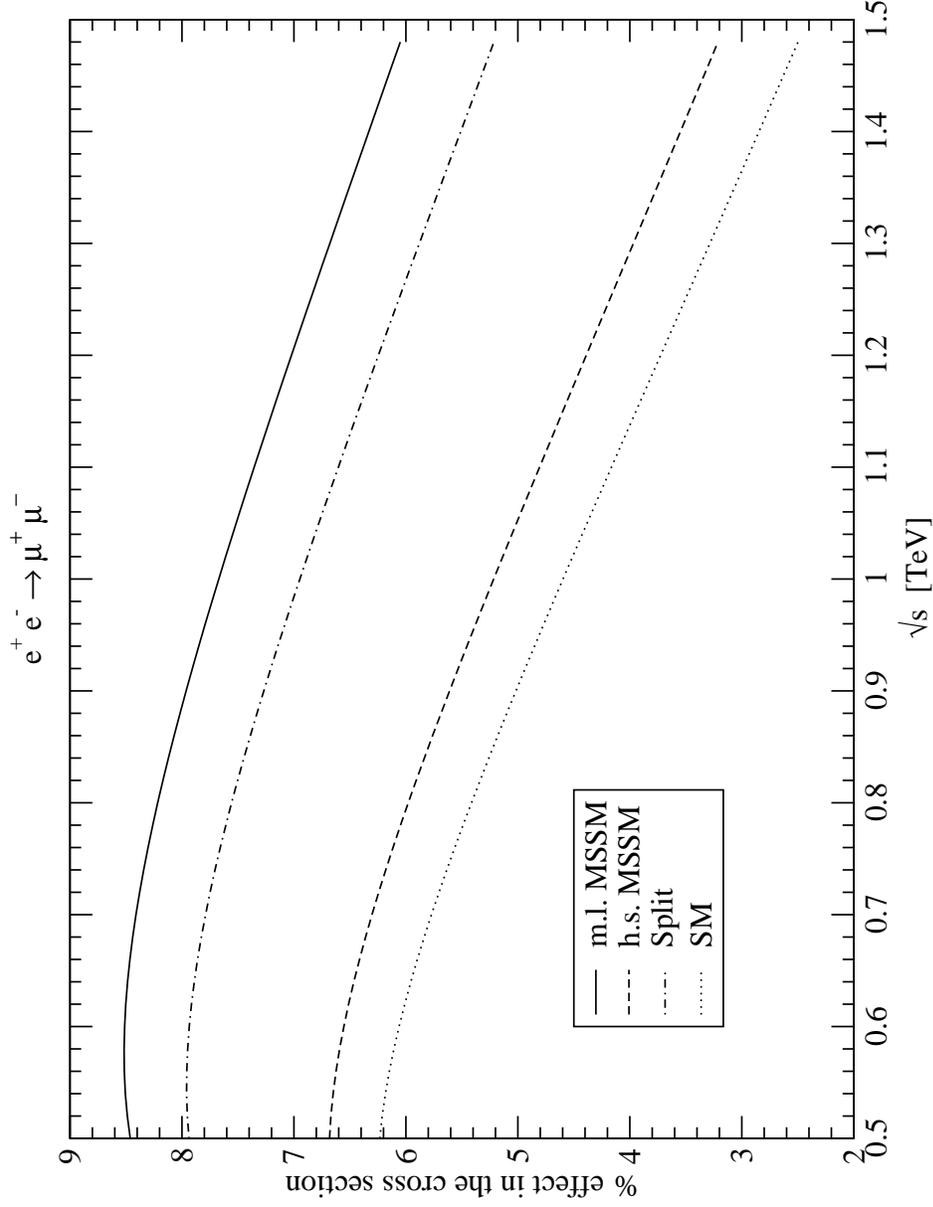}
\vspace{1.5cm}
\caption{
Relative percentual effect in the cross section at partonic level 
for the process $e^+e^-\to \mu^+\mu^-$
in various scenarios.
There is no $\tan\beta$ dependence both in ``m.l. MSSM'' and in ``h.s. MSSM''.
}
\label{fig9}
\end{figure}

\newpage

\begin{figure}
\centering
\epsfig{file=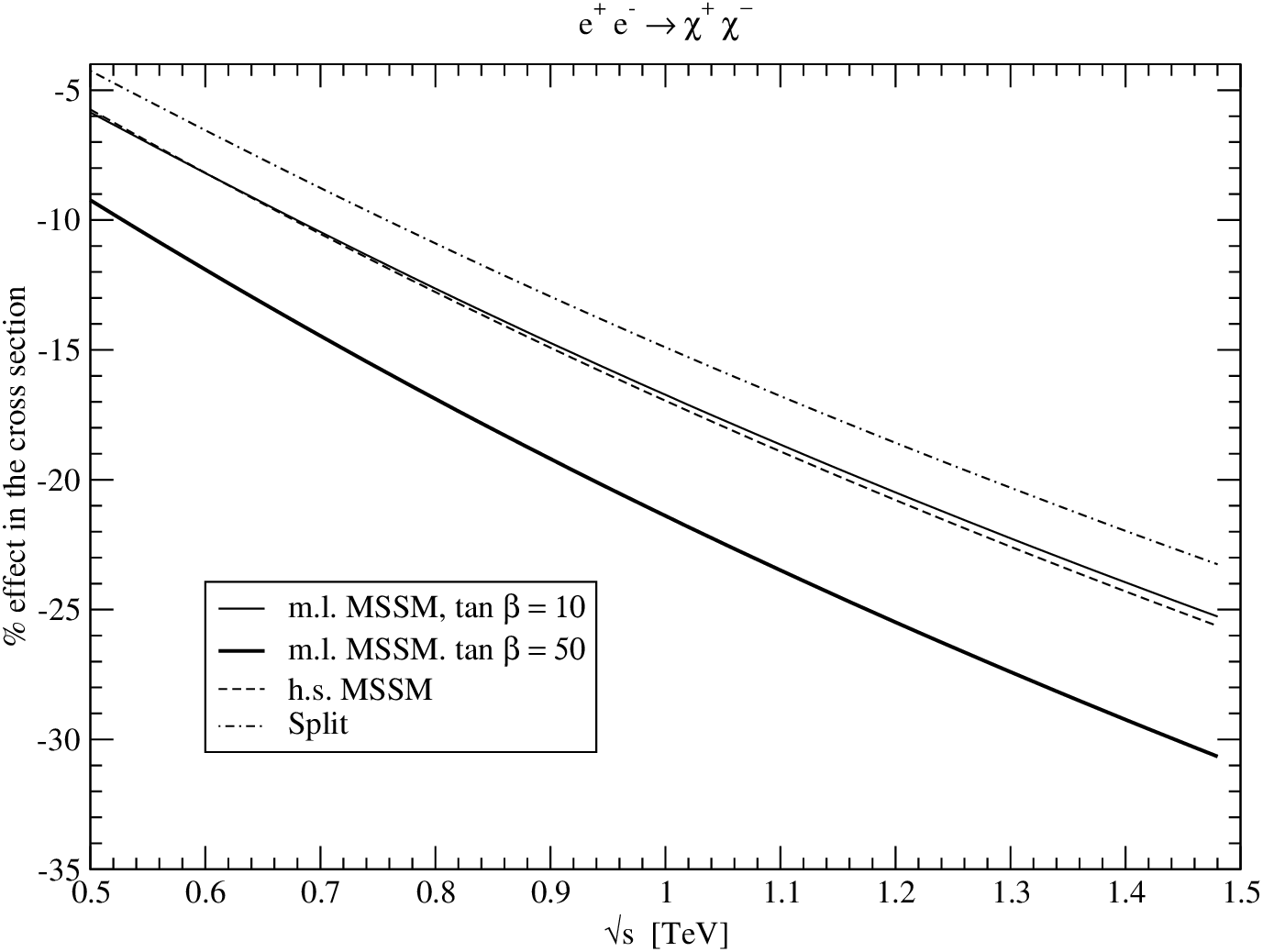,width=16cm,angle=90}
\vspace{1.5cm}
\caption{
Relative percentual effect in the cross section at partonic level 
for the process $e^+e^-\to \chi^+\chi^-$
in various scenarios.
Notice that there is no $\tan\beta$ dependence in the model ``h.s. MSSM''.
}
\label{fig10}
\end{figure}

\end{document}